\documentclass[prl,twocolumn,showpacs,floatfix,amsmath,amssymb,superscriptaddress]{revtex4-1}
\usepackage{amsfonts}
\usepackage{stmaryrd}
\usepackage{bbm}
\usepackage{mathrsfs}
\usepackage{tipa}
\usepackage{amssymb}
\usepackage{txfonts}
\usepackage{graphicx}
\usepackage{dcolumn}
\usepackage{epstopdf}
\usepackage[colorlinks,linkcolor=blue,urlcolor=blue,citecolor=blue]{hyperref}
\usepackage{multirow}
\usepackage{subfigure}
\usepackage{url}

\begin{document}

\title{Optical study of three-dimensional Weyl semimetal Mn$_3$Sn} %force line break\\

\author{L. Y. Cao}
\affiliation{Center for Advanced Quantum Studies and Department of Physics, Beijing Normal University, Beijing 100875, China}

\author{Z. A. Xu}
\affiliation{School of Physical Science and Technology, ShanghaiTech University, Shanghai 201210, China}

\author{B. X. Gao}
\affiliation{Center for Advanced Quantum Studies and Department of Physics, Beijing Normal University, Beijing 100875, China}

\author{L. Wang}
\affiliation{Center for Advanced Quantum Studies and Department of Physics, Beijing Normal University, Beijing 100875, China}

\author{X. T. Zhang}
\affiliation{Center for Advanced Quantum Studies and Department of Physics, Beijing Normal University, Beijing 100875, China}

\author{X. Y. Zhang}
\affiliation{Center for Advanced Quantum Studies and Department of Physics, Beijing Normal University, Beijing 100875, China}

\author{Y. F. Guo}
\affiliation{School of Physical Science and Technology, ShanghaiTech University, Shanghai 201210, China}
\affiliation{ShanghaiTech Laboratory for Topological Physics, Shanghai 201210, China}

\author{R. Y. Chen}
\affiliation{Center for Advanced Quantum Studies and Department of Physics, Beijing Normal University, Beijing 100875, China}

\begin{abstract}

Three-dimensional (3D) Weyl semimetal Mn$_3$Sn has attracted tremendous attention due to its great application potential. However, the complex magnetic structures at different temperature intervals make it extremely difficult to unravel the underlying electronic structures of Mn$_3$Sn.
Here, we perform temperature-dependent optical spectroscopy measurements on single crystalline Mn$_3$Sn to investigate its charge dynamics. We find that both of the optical reflectivity $R(\omega)$ and conductivity $\sigma_1(\omega)$ evolve very smoothly across the magnetic phase transition at $T_M$ = 285 K, where the giant anomalous Hall effect (AHE) at room temperature drops significantly.
Furthermore, two linearly increasing segments of $\sigma_1(\omega)$ are observed in the whole temperature range  from 300 K to 10 K, indicating that the existence of Weyl fermions is very robust against the magnetic phase transition. In addition, the Weyl points closest to the Fermi level $E_F$ are identified to be located about 101 meV away from $E_F$ at 10 K, and the associated Fermi velocity is about 2.55 $\times 10^7$ cm/s. Our results reveal that the phase transition at $T_M$ only generates subtle modification to the band structure, which helps to further uncover the mechanism of the dramatic change of AHE in Mn$_3$Sn.

%indicating the presence of 3D Weyl femions. The Fermi velocities of 3D Weyl fermions are $v_{F1} =  and $v_{F2} = 9.56 \times 10^6$ cm/s at 10 K, which is associated with the different interband transitions close to Weyl points. The Fermi velocity of Weyl fermions corresponding to the low-energy linear $\sigma_1(\omega)$ decreases upon cooling, while that of the high-energy linear $\sigma_1(\omega)$ is temperature-independent. We obtain that the potential of the Fermi level relative to the Weyl point are about 101 meV and 254 meV.

%Many exotic phenomena have been found due to non-zero Berry curvature, whereas the physical properties affected by strongly electron correlation remain to be explored.
%However, many physical properties are still unknown due to strongly electron correlation effect. The density of free carriers derived from the real part of the optical conductivity increases continuously with the temperature decreasing, which confirms Mn$_3$Sn is a good metal.

\end{abstract}

\maketitle
\section{introduction}

Non-trivial topological states have been a hot topic in condensed matter physics ever since its discovery, which is of great significance for both of  fundamental physics and potential applications. Among the big family of topological states, including topological insulators\cite{RevModPhys.82.3045,RevModPhys.83.1057,doi:10.1126/science.1133734}, topological superconductors\cite{RevModPhys.83.1057,doi:10.1146/annurev-conmatphys-030212-184337}, topological semimetals\cite{PhysRevB.83.205101,doi:10.1126/science.aaa9297,Weng_2016} and so on, Weyl semimetals (WSMs) are of special interests due to its unique Berry curvature distribution. It is well known that Berry curvature acts like magnetic monoples at the Weyl points in the momentum space, which leads to a lot of exotic phenomena, such as anomalous Hall effect (AHE)\cite{doi:10.1146/annurev-conmatphys-031016-025458,Burkov_2015}, and anomalous Nernst effect (ANE)\cite{PhysRevB.93.035116,PhysRevB.96.155138}, etc. Especially, AHE was initially believed to exist only in ferromagnetic materials and its magnitude is closely related to the net magnetizations. The development of the so called ``intrinsic mechanism'' associated with the large Berry curvature greatly expands the possible candidates of AHE materials.

Specifically, the compound Mn$_3$Sn is the first antiferromagnetic (AFM) material with a vanishingly small net magnetization that exhibits giant AHE at room temperature\cite{RN436,doi:10.1063/1.5021133,PhysRevB.101.144422}. Meanwhile, a series of intriguing phenomena like ANE\cite{PhysRevLett.119.056601,RN437,PhysRevLett.119.056601}, anomalous thermal Hall effect\cite{PhysRevLett.119.056601}, magneto-optical Kerr effect\cite{RN446}, and topological Hall effect\cite{doi:10.1063/1.5119838,PhysRevB.99.094430} are observed in Mn$_3$Sn as well. Additionally, the small net magnetization in Mn$_3$Sn enables easy manipulation of the anomalous effects via merely weak magnetic field or electrical current\cite{RN436,RN488}, making the compound a very promising candidate for next-generation technology, such as high-density and ultrafast-operation spintronic devices. It is well established that these unexceptionable properties come from the fact that Mn$_3$Sn is a time reversal symmetry breaking Weyl semimetal, which has been verified by the combination of theoretical calculations\cite{Yang_2017} and observation of chiral anomaly in transport measurements\cite{RN68}.

Mn$_3$Sn crystallize is a hexagonal Ni$_3$Sn-type lattice structure with a space group $P6_3/mmc$\cite{PJBrown1990}. Mn atoms constitute kagome layers in the $ab$ plane, and these kagome planes are stacked along the $c$ axis. Below $T_N$ = 420 K, Mn$_3$Sn enters the AFM phase, where the magnetic moments of Mn atoms are reported to be arranged in the $ab$ plane and rotated 120$^{\circ}$ with a negative vector chirality. Density functional theory (DFT) calculations suggest there are dozens of three-dimensional (3D) Weyl points near the Fermi level under this magnetic configuration\cite{Yang_2017}, which can perfectly account for the appearance of giant AHE and ANE. However, the giant AHE may experience dramatic changes and almost disappear at lower temperatures, the underlying mechanism of which is believed to be closely related to the magnetic phase transitions.

When temperature decreases, the magnetic structure experiences different phase transitions
%The observation of chiral anomaly further verifies the existence of Weyl fermions in Mn$_3$Sn.
depending on the growth condition and/or the compositions of the elements\cite{KREN1975226}.
Magnetic phase transition from the chiral triangular AFM state to an incommensurate spiral magnetic structure is reported to take place at around $T_M$ $\sim$ 275 K in Mn$_3$Sn grown by flux method\cite{doi:10.1063/1.5119838,doi:10.1063/1.5021133}. Simultaneously, the large AHE disappears abruptly below $T_M$. A similar AFM to incommensurate transition is revealed in samples grown using the arc-melting method, only that it occurs slowly from 250 K to 190 K\cite{PhysRevB.101.144422}. Meanwhile, the AHE gradually fades away in the same manner. In Czochralski-method grown samples with a bit excessive Mn element\cite{RN436,RN437}, however, no magnetic phase transitions are observed, in which case the AHE can persist down to $T_g\sim$ 50 K, below which a spin-glass phase is generally demonstrated\cite{PhysRevB.101.144422,doi:10.1063/1.5021133}.
%In any cases, the variations of AHE always keep in step with the magnetic structure transitions, which is reported to be crucial to the positions and numbers of the Weyl points.

At the current state, there are still full of questions concerning the complex magnetic structures at different temperature intervals, let alone the corresponding electronic structures which is crucial to the variation of AHE.
To further reveal the underlying physics of Mn$_3$Sn, especially below the magnetic phase transition temperatures, we perform optical spectroscopy measurements on single crystalline Mn$_3$Sn compounds from 300 K to 10 K. We find that the real part of optical conductivity shows linear increasing features as a function of frequency at two different energy regions, which is usually considered as an evidence of 3D linear dispersive band structures, in agreement with the consensus of Mn$_3$Sn being a Weyl semimetal. More importantly, this character persists from room temperature down to 10 K, indicating the robustness of the Weyl fermions against the magnetic phase transition at $
T_M$. %We evaluate the momentum-relaxation is fast. ($E_F$

%in the range of 1700 cm$^{-1}$ to 3500 cm$^{-1}$ and 6000 cm$^{-1}$ to 12 000 cm$^{-1}$

%We evaluate the momentum-relaxation is fast, according to the momentum-relaxation length $l_{mr}$ is at the nanometer scale at 10 K.

\section{experiment}

Single crystals of Mn$_3$Sn were grown by using the flux method. Mn pieces (99.98\%), Sn grains (99.999\%) and Bi grains (99.999\%) were mixed with an atomic ratio of 3.1:1:7 and placed into an alumina crucible. The crucible was sealed in a fused silica ampoule in vacuum. The ampoule was slowly heated up to 1000 $^{\circ}$C in a furnace at a rate of 100 $^{\circ}$C/h. After maintaining at this temperature for 15 hours, the ampoule was cooled down to 800 $^{\circ}$C over 2 hrs and then slowly cooled to 300 $^{\circ}$C at the rate of 3 $^{\circ}$C/h. At 300 $^{\circ}$C, the silica ampoule was taken out from the furnace and immediately centrifuged to separate the crystals from the Bi flux. The temperature-dependent resistivity $\rho(T)$ and the magnetic susceptibility $\chi_{ab}(T)$ up to 350 K were measured in a Quantum Design physical property measurement system.

Infrared spectroscopy studies are performed with a Bruker IFS 80 V in the frequency range from 30 to 42 000 cm$^{-1}$. An $in situ$ gold and aluminium overcoating technique is used to get the reflectivity $R(\omega)$. The real part of the optical conductivity $\sigma_1(\omega)$ is obtained by the Kramers-Kronig transformation of $R(\omega)$. The Hagen-Rubens relation is used for the low-frequency extrapolation of $R(\omega)$. We employ the x-ray atomic scattering functions in the high-energy side extrapolation\cite{PhysRevB.91.035123}.

\section{results and discussion}

\begin{figure}[htb]
\centering
\includegraphics[width=9cm]{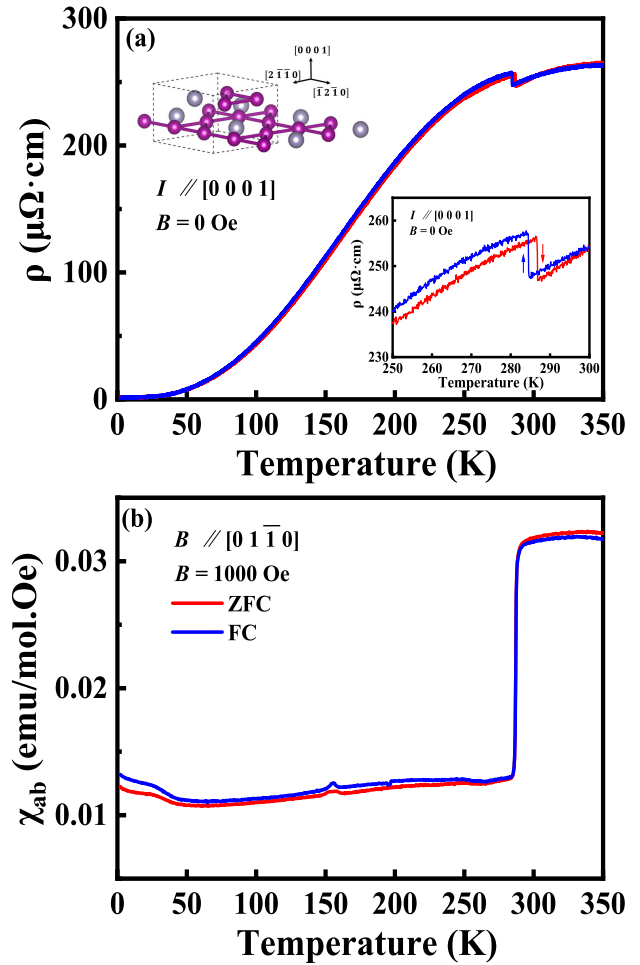}
\caption{(a) The resistivity as a function of temperature of Mn$_3$Sn. The inset displays the crystal structure of Mn$_3$Sn. (b) The magnetic susceptibility as a function of temperature of Mn$_3$Sn, which is measured with $\emph{B =}$ 1000 Oe along $\emph{B $\parallel$}$ [0 1 $\overline{1}$ 0] in both ZFC and FC modes. }
\label{Fig:1}
\end{figure}

Figure \ref{Fig:1}(a) shows the resistivity of Mn$_3$Sn as a function of temperature. When the current is applied along with the [0 0 0 1] direction, the resistivity $\rho(T)$ shows an overall metallic behavior. A sudden jump is observed at $T_M$ = 285 K, signaling the transition between the non-collinear chiral AFM state and the incommensurate magnetic state. At the same time, a hysteresis behavior of $\rho(T)$ appears during warming and cooling, indicating that this is a first-order phase transition, which is basically consistent with previous results of stoichiometrical samples\cite{doi:10.1063/1.5021133,doi:10.1063/1.5119838}. The temperature-dependent magnetic susceptibility with zero-field-cooling (ZFC) and field-cooling (FC) modes at $\emph{B =}$ 1000 Oe for $\emph{B $\parallel$}$ [0 1 $\overline{1}$ 0] of Mn$_3$Sn is plotted in Fig.~\ref{Fig:1}(b). A significant change is found at around 285 K, which verifies the occurrence of the magnetic phase transition at $T_M$. Notably, a magnetic helical structure reconstruction transition at around 200 K is reported by Sung et al.\cite{doi:10.1063/1.5021133}, which is absent neither in our measurement, nor in the measurements by P J Brown et al. and Kuroda et al.\cite{PJBrown1990,TOMIYOSHI19861001,RN436}. This further demonstrates the complexity of the magnetic structures of Mn$_3$Sn samples, affected by the growth methods and/or the ratio of the elements. Below $T_g$ $\sim$ 50 K, $\chi_{ab}(T)$ rises with temperature decreasing and a hysteresis in $\chi_{ab}(T)$ is observed between ZFC and FC modes, which is agreement with the spin-glass state as suggested previously\cite{doi:10.1063/1.5021133}.

%which may be due to the inconspicuous reconstruction of the magnetic moment of the measured sample, or it may be related to the sample quality.

\begin{figure}[t]
\centering

  \includegraphics[width=9cm]{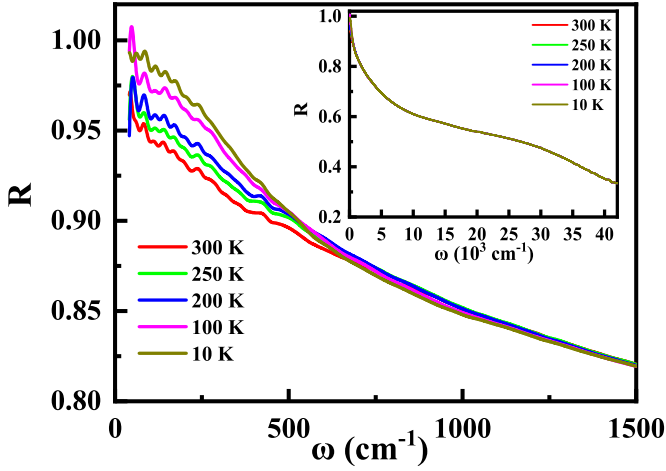}

  \caption{The frequency dependent reflectivity $R(\omega)$ of Mn$_3$Sn. The inset shows $R(\omega)$ in an expanded range up to 42 000 cm$^{-1}$.}
  \label{Fig:2}
\end{figure}

The reflectivity spectra $R(\omega)$ up to 1500 cm$^{-1}$ of Mn$_3$Sn is shown in the main panel of Fig.~\ref{Fig:2}, while the expanded range up to 42 000 cm$^{-1}$ is plotted in its inset.  In the low-frequency range, $R(\omega)$ increases with frequency decreasing and tends to unit at zero frequency, indicating a metallic nature. Additionally, the low-frequency reflectivity increases upon cooling from 300 K to 10 K, which is in accordance with the result of the transport measurement. Above 1200 cm$^{-1}$, $R(\omega)$ at different temperatures almost overlaps with each other, revealing the charge dynamics are mainly affected by temperature in a very narrow energy window. Surprisingly, $R(\omega)$ evolves very smoothly across $T_M$, in spite of the sudden jump of $\rho(T)$ at $T_M$, suggesting the electronic structure is actually quite stable against this magnetic phase transition.

\begin{figure}[htbp]
  \centering

   \includegraphics[width=9cm]{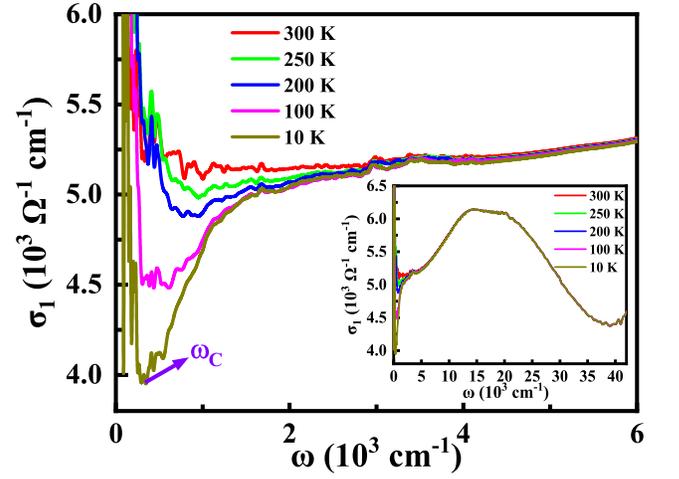}

  \caption{The real part of optical conductivity $\sigma_1(\omega)$ of Mn$_3$Sn. The inset shows $\sigma_1(\omega)$ in an expanded range up to 42 000 cm$^{-1}$. The violet arrow indicates cut-off frequency at 10 K.}
  \label{Fig:3}
\end{figure}

In order to further investigate the charge dynamic responses, we derive the real part of optical conductivity $\sigma_1(\omega)$ of Mn$_3$Sn. Figure~\ref{Fig:3} displays $\sigma_1(\omega)$ below 6000 cm$^{-1}$ at several selected temperatures in its main panel, and the inset shows $\sigma_1(\omega)$ in a large energy scale ranging up to 42 000 cm$^{-1}$. The direct-current conductivity $\sigma_1(\omega \rightarrow 0)$ increases with temperature decreasing, which is roughly consistent with the behavior of 1/$\rho(T)$. Approaching zero frequency, $\sigma_1(\omega)$ increases rapidly with frequency decreasing, which can be considered as a peak centered at $\omega = 0$. Such characters are usually attributed to intraband transitions of free carriers, which can be described by the Drude model
\begin{equation}
\label{equ:1}
\sigma_{1-Drude}(\omega)=\frac{{\omega_P}^2\tau}{4\pi}\frac{1}{1+{\omega}^2{\tau}^2}.
\end{equation}
In the above equation $\tau$ is the relaxation time of free carriers, and $\omega_P$ is the plasma frequency which is related to the carrier density $n$ and effective mass $m^{*}$ by $\omega_P=\sqrt{4\pi n e^2/m^{*}}$.
Obviously, the Drude component of $\sigma_1(\omega)$ can be resolved at all the measurement temperatures, which is in correspondence with the metallic nature of Mn$_3$Sn. The width at half maximum of the Drude peak indicates the scattering rate of free carriers $\gamma$ = 1/$\tau$, which decreases upon cooling, as can be seen in Fig.~\ref{Fig:3}, agreeing well with the behavior of $\rho(T)$.

\begin{figure}[htbp]
  \centering

   \includegraphics[width=9cm]{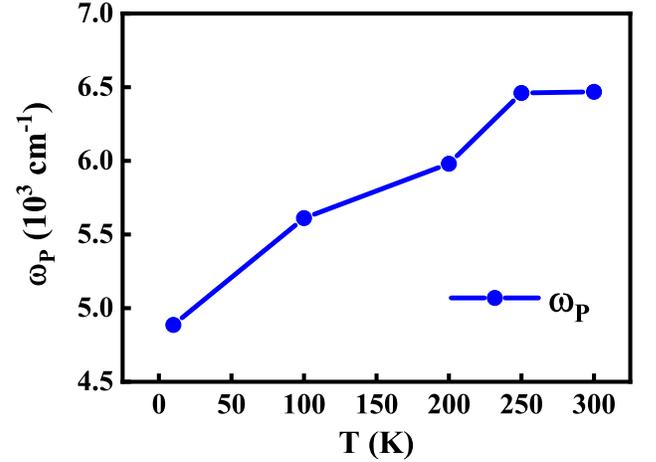}

  \caption{Temperature-dependent plasma frequency $\omega_P$ of Mn$_3$Sn.}

  \label{Fig:4}
\end{figure}

The plasma frequency of $\omega_P$ could be obtained from the integration of the Drude spectral weight by $\omega_P^2 = 8 \int_{0}^{\infty} \sigma_{1-Drude}(\omega) \ d \omega $, which should include the contributions of all intraband transitions, while excludes any interband ones. However, since intraband and interband transitions overlap with each other in a wide energy range, we use  $\omega_P^2 = 8 \int_{0}^{\omega_c} \sigma_{1}(\omega) \ d \omega $ to estimate the plasma frequency, in which $\omega_c$ is the cutoff frequency where $\sigma_1(\omega)$ reaches a minimum value. As indicated in Fig.~\ref{Fig:3}, the value of $\omega_c$ is identified to be 345, 612, 776, 952, and 1001 cm$^{-1}$ in a sequence corresponding to increasing temperatures. The obtained temperature-dependent $\omega_P$ is plotted in Fig.~\ref{Fig:4}. We find that $\omega_P$ stays almost unchanged across the magnetic phase transition at $T_M$, then it drops mildly from 6461 cm$^{-1}$ at 250 K to 4886 cm$^{-1}$ at 10 K. Notably, a previous terahertz conductivity study on thin film Mn$_3$Sn observes a very similar behavior of the plasma frequency\cite{RN338}.

Since $\omega_P^2$ is proportional to the carrier density $n$ and inversely proportional to the effective mass $m^\ast$, the decreasing of $\omega_P$ upon cooling could be caused by reduction of carrier density and/or enhancement of the effective mass. Bing Cheng et al. proposed that the breaking of translation symmetry below $T_M$ might lead to partial gapping of the Fermi surfaces\cite{RN338}. Very recently, two independent neutron scattering experiments suggest either a spin density wave gap\cite{wang2023flat} or a charge density wave gap\cite{chen2023cdw} develops below $T_M$. These scenarios seem to be able to explain the continuous loss of carrier density as temperature lowers. However, such gap opening behaviors, which usually give rise to Lorentz-type peaks centered at the gap energy in $\sigma_1(\omega)$, can not be unambiguously observed in our results.
Since Mn$_3$Sn is believed to be a Weyl semimetal, the slightly shift of chemical potential relative to Weyl points may induce a sizable modulation of the carrier density as well\cite{PhysRevB.89.245121}. Therefore, it is probably that the chemical potential moves closer to the Weyl point as temperature decreases, which induces the reduction of carrier density $n$.
%If we assume the effective mass remains constant with cooling, the reduction of $\omega_P$ suggests a continuous loss of the density of free carriers.

On the other hand, ARPES measurement together with DFT calculations demonstrates that the electronic correlation in Mn$_3$Sn is quite strong, which renormalizes the band structure by a factor of 5\cite{RN68}, corresponding to a large effective mass.
%There is also a possibility that the correlation effect grows even stronger at lower temperatures, which can account for the decreasing of $\omega_P$ as well.
According to former experimental and theoretical results, the carrier density $n$ is estimated to be roughly around 1.5 $\times 10^{22}$cm$^{-3}$ \cite{RN436, PhysRevLett.119.056601, RN472, 10.21468/SciPostPhys.5.6.063}.
If assuming $m^{*}$ is the statice electron mass in our experiment, $n$ is identified to be 4.66 $\times 10^{20}$ cm$^{-3}$ at 300 K and 2.66 $\times 10^{20}$ cm$^{-3}$ at 10 K via $\omega_P^2=\frac{4\pi n e^2}{m^\ast}$, which are almost two orders of magnitude smaller than previous reports. Reversely, if we adopt the carrier density of $n=$ 1.5 $\times 10^{22}$ cm$^{-3}$, we could yield ${m^\ast}/{m_e} \approx 32$, indicating extremely strong electron correlation effect in Mn$_3$Sn.

\begin{figure}[t]
\centering

  \includegraphics[width=9cm]{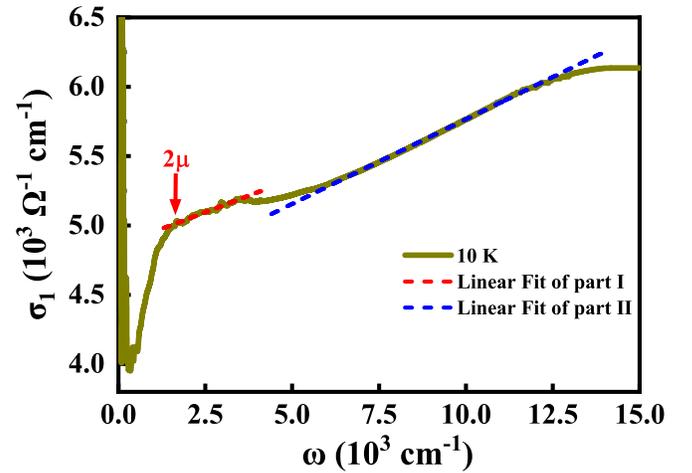}

  \caption{The optical conductivity below 15 000 cm$^{-1}$ at 10 K. The red and blue short dashed lines present the linear fitting results of $\sigma_1(\omega)$. The step-like feature is indicated by the red arrow.}
  \label{Fig:5}
\end{figure}

Above the Drude component, two segments of linear growing $\sigma_1(\omega)$ could be observed. To get a better vision, we replot the optical conductivity at 10 K up to 15 000 cm$^{-1}$ in Fig.~\ref{Fig:5}. It is clearly seen that $\sigma_1(\omega)$ rapidly increases until about 1500 cm$^{-1}$, where a segment of linear increasing optical conductivity sets in, as indicated by the red dashed line in Fig.~\ref{Fig:5}. As frequency increases, another segment of linear $\sigma_1(\omega)$ is observed around 6000 - 12 500 cm$^{-1}$, as marked by the blue dashed line.
Subsequently, the optical conductivity decreases gradually with the frequency increasing. It is worth noting that the two linear parts of $\sigma_1(\omega)$ cannot be reproduced by the Lorentz model, which is normally used to describe interband transitions. Instead, it is analogy to the charge dynamics of a great deal of topological semimetals with linear dispersive band structures, such as 3D Dirac semimetal ZrTe$_5$\cite{Chen2015}, Weyl semimetal TaAs\cite{PhysRevB.93.121110}, nodal line semimetal NbAs$_2$\cite{RN227}, and so on. The optical transitions of the above listed noninteracting electron systems consisting of two symmetric bands touching each other at the Fermi energy have been extensively studied. Their optical conductivity has power-law frequency dependence with $\sigma_1(\omega) \propto (\frac{\hbar\omega}{2})^\frac{d-2}{z}$, where $d$ is the dimension of the system and $z$ refers to the power-law term of the band dispersion\cite{PhysRevB.87.125425}.
%The above formula has been applied to a variety of optical researches of the topological semimetals. For example, the optical conductivity is independent with frequency for 2D topological semimetal ZrSiS\cite{PhysRevLett.119.187401} and graphere\cite{PhysRevLett.101.196405} ($d = 2$ and $z = 1$). Unlike the flat optical conductivity, the real part of the optical conductivity is found to increase linearly with frequency due to linear energy band dispersion in 3D Dirac semimetal ZrTe$_5$\cite{Chen2015}, 3D Weyl semimetal NbP\cite{PhysRevB.98.195203} and YbMnBi$_2$\cite{PAL201864} ($d = 3$ and $z = 1$).
For the strongly correlated 3D material Mn$_3$Sn, the linear frequency-dependent optical conductivity not only confirms the presence of Weyl fermions at room temperature, but also provides strong evidence for its persisting existence down to the lowest temperature.

%with different slopes from 300 K to 10 K, ranging from about 1700 cm$^{-1}$ to 3500 cm$^{-1}$ and 6000 cm$^{-1}$ to 12 000 cm$^{-1}$, respectively. The linear optical conductivity while
%In order to make it clearly to see...%which appears as a peak centered on the resonant frequency $\omega_L$.

Furthermore, the Fermi velocity of the Weyl fermions could be estimated via
\begin{equation}\label{Eq:2}
\sigma_1(\omega) = \frac{{e^2}{N_W}}{12h} \frac{\omega}{v_F} \theta(\omega - 2 \lvert \mu \rvert),
\end{equation}
where $N_W$ is the number of Weyl points, $h$ is the Planck constant, $v_F$ is the Fermi velocity, $\theta(x)$ is Heaviside step function, and $\mu$ is the potential of the Fermi level relative to the Weyl point\cite{PhysRevLett.108.046602,PhysRevB.93.085426}.
This equation describes the interband transitions purely contributed by massless Dirac or Weyl fermions, therefore $\sigma_1(\omega)$ is supposed to be extrapolated to zero at zero frequency, when the Fermi level lies exactly at the Dirac or Weyl points. If $\mu$ has a finite value, then the interband transition will be terminated when $\omega \leq 2 \lvert \mu \rvert$ due to the Pauli blocking effect, resulting in a step-like feature of the optical conductivity prior to the emergence of linear growth. For Mn$_3$Sn, both of the two linear segments have a positive intercept at zero frequency, possibly ascribed to interband transitions between trivial bands. Additionally, a very clear step-like feature is observed for the lower energy linear part, signaling that the Weyl points are located away from the Fermi energy.
%Although it is hard to validate such features for the higher energy linear part, which is again most probably blurred out by the contribution of trivial bands, it starts to exhibit itself somewhere around 5000 cm$^{-1}$.
With temperature increasing, this step feature is gradually blurred out by the free electron excitations due to increasing scattering rate.

In order to pin down the positions of Weyl points, we yield the first-order derivative of $\sigma_1(\omega)$ and take the inflection point as the value of $2 \lvert \mu \rvert$, which is identified to be about 1630 cm$^{-1}$ (202 meV) at 10 K, as marked by the red arrow in Fig.~\ref{Fig:5}.
%which indicates the potential of Fermi level with respect to the Weyl point are 101 meV and 254 meV.
DFT calculations have predicted that Mn$_3$Sn processes six pairs of Weyl points considering a noncollinear AFM ground state, which can be classified into three groups according to their positions. W$_1$, W$_2$ and W$_3$ are located relative to the Fermi energy of 86 meV, 158 meV, and 493 meV, where each type of Weyl points has four copies\cite{Yang_2017}. Following this configuration, the low-energy linear part of $\sigma_1(\omega)$ is likely associated with the interband transitions close to W$_1$. According to Eq.~\ref{Eq:2}, we can obtain that the corresponding Fermi velocity $v_{F1}$ reduces from 5.41 $\times$ 10$^7$ cm/s at 300 K to 2.55 $\times$ 10$^7$ cm/s at 10 K, taking $N_W$ = 4.
Since the higher energy linear part of $\sigma_1(\omega)$ persist all the way to 12 500 cm$^{-1}$($\sim$ 1.55 eV), we estimated $v_{F2} = 5.98 \times 10^7$ cm/s using $N_W$ = 12, which remains unchanged in the whole temperature range.

%Recently, Li et al. proposes a doubly-degenerate nodal line without SOC exists in Mn$_3$Sn, referred to as the Weyl nodal line, along the $K-H$ axis in the Brillouin zone. In addition, the tiny hybridization gaps are induced considering SOC\cite{RN472}. If we consider the band structure with Weyl nodal line, the linear optical conductivity remains, while the value of $2 \lvert \mu \rvert $ represents the sum of the double potential of the Fermi level relative to the minimum of conduction band and energy scale of the band gap (2$\Delta$) caused by SOC. Therefore, further investigations are still required to reveal the underlying physics.

It is worth to remark that charge dynamics of Mn$_3$Sn exhibits nearly no alteration across $T_M$, except for a slight reduction of the scattering rate and the Fermi velocity of parts of the Weyl fermions. As elaborated above, both of the magnetic structure and AHE experience significant change at this temperature, and it is generally believed the underlying electronic structure is strongly modified as well.
%t is worth to remark that the above estimation is based on an assumption that the band structure, especially the number and positions of the Weyl points do not vary with temperature changing. In reality, however, the band structure is supposed to be very sensitive to the magnetic structures, which experience multiple transitions upon cooling.
For example, Pyeongjae Park et al. propose that the Weyl points in Mn$_3$Sn might be eliminated in the helical phase below $T_M$\cite{RN489}, which is responsible to the sudden reduction of AHE. However, our results demonstrate that the band structure is actually quite stable against the magnetic phase transitions. Especially, the Weyl fermions are not only robust across $T_M$, but even down to the lowest temperature, although the position of Weyl points relative to the Fermi energy might be shifted. Therefore, the enormous change of AHE is more likely to be associated with the subtle distribution of the Weyl points.

\section{conclusion}

In conclusion, we have investigated optical responses of Mn$_3$Sn single crystals. Metallic behaviors are clearly manifested from 300 K to 10 K. The corresponding carrier density stays almost unchanged across the magnetic phase transition at $T_M$ = 285 K, then decreases slightly upon cooling, which might be caused by the shifting of Weyl points relative to the Fermi level. The effective mass is estimated to be as large as ${m^\ast}/{m_e} \approx 32$, which is consistent with the strongly electron correlation effect in Mn$_3$Sn. More importantly, two linearly increasing parts of the real part of the optical conductivity are observed in the whole temperature range, from 1630 cm$^{-1}$ to 3550 cm$^{-1}$ and 6000 cm$^{-1}$ to 12 500 cm$^{-1}$, respectively. This character provides strong evidence of the continuous presence of 3D Weyl fermions across $T_M$. The slope of the lower energy one increases with cooling, while the slope of the higher energy one stays unchanged. The corresponding Fermi velocities are $v_{F1} = 2.55 \times 10^7$ cm/s and $v_{F2} = 5.98 \times 10^7$ cm/s at 10 K. In addition, we deduce the energy difference between the Fermi level and the closest Weyl points is about 101 meV. Our results not only provide new information on the electronic structure of Mn$_3$Sn, but also inspire further researches exploring the mechanism of the significant change of AHE.

\begin{center}
\small{\textbf{ACKNOWLEDGMENTS}}
\end{center}

R. Y. Chen acknowledges the support by the National Natural Science Foundation of China (Grant No. 12074042), the National Key Projects for Research and Development of China (Grant No. 2021YFA1400400, 2016YFA0302300), the Young Scientists Fund of the National Natural Science Foundation of China (Grant No. 11704033).
Y. F. Guo acknowledges the support by acknowledges the open projects from State Key Laboratory of Functional Materials for Informatics (Grant No. SKL2022), CAS, and the Double First-Class Initiative Fund of ShanghaiTech University.

%Y.F.G. acknowledges the support by acknowledges the open projects from State Key Laboratory of Functional Materials for Informatics (Grant No. SKL2022), CAS, and the Double First-Class Initiative Fund of ShanghaiTech University.

\bibliographystyle{apsrev4-1}
  \bibliography{ReferenceMn3Sn}

\end{document}